\documentclass[12pt, draftclsnofoot, onecolumn]{IEEEtran}

\usepackage{graphicx}  %[dvips]���ܼӣ�ֱ��ת��pdf �ļ���
\usepackage{epstopdf}
\usepackage{epsfig}
\usepackage{subfigure}
\usepackage[cmex10]{amsmath}
\usepackage{amssymb}   % �������塢���Ŷ����ˡ� ��
\usepackage{amsmath}
\usepackage{amsfonts}
\usepackage{cite}
\usepackage{color}

\newtheorem{myTheo}{Theorem}

\newtheorem{myOb}{Observation}
\newtheorem{myProp}{Proposition}

\begin{document}

\title{Rethinking Uplink Hybrid Processing: When is Pure Analog Processing Suggested?}

\author{Jingbo Du,~\IEEEmembership{Student Member,~IEEE,}
		Wei Xu,~\IEEEmembership{Senior Member,~IEEE,}
		\\Bin Sheng,~\IEEEmembership{Member,~IEEE,}
		and Chunming Zhao,~\IEEEmembership{Member,~IEEE,}

\thanks{%Copyright (c) 2015 IEEE. Personal use of this material is permitted. However, permission to use this material for any other purposes must be obtained from the IEEE by sending a request to pubs-permissions@ieee.org.

%Manuscript received June 21, 2018; revised December 03, 2018 and February 25, 2019; accepted March 04, 2019.

%This work was supported in part by the National Science and Technology Projects of China under Grant 2018ZX03001002, the NSFC under 61871109, the Hong Kong, Macao and Taiwan Science \& Technology Cooperation Program of China under 2016YFE0123100, the Six Talent Peaks project in Jiangsu Province under GDZB-005, and the China Scholarship Council under Grant 201706090067. 

J. Du is with the National Mobile Communications Research Laboratory, Southeast University, Nanjing 210096, China, and also with the Institute of Information and Communication Technologies, Electronics and Applied Mathematics, Universit\'{e} catholique de Louvain, 1348 Louvain-la-Neuve, Belgium (email: 230159371@seu.edu.cn). Wei Xu, Bin Sheng and Chunming Zhao are with the National Mobile Communications Research Laboratory, Southeast University, Nanjing 210096, China (email: \{wxu, sbdtt, cmzhao\}@seu.edu.cn). \emph{(Corresponding author: Wei Xu.)}}
}

%\markboth{IEEE Transactions on Vehicular Technology,~Vol.~XX, No.~XX, XXX~2019}
%{Du \MakeLowercase{\textit{et al.}}: Rethinking Uplink Hybrid Processing: When is Pure Analog Processing Suggested?}

\maketitle

\begin{abstract}
In this correspondence, we analytically characterize the benefit of digital processing in uplink massive multiple-input multiple-output (MIMO) with sub-connected hybrid architecture. By assuming that the number of radio frequency (RF) chains is equal to that of users, we characterize achievable rates of both pure analog detection and hybrid detection under the i.i.d. Rayleigh fading channel model. From the derived expressions, we discover that the analog processing can outperform the hybrid processing using the maximal ratio combining (MRC) or zero-forcing (ZF) criterion in cases under some engineering assumptions. Performance comparison of the schemes are presented under tests with various numbers of users and numbers of antennas at the base station.
\end{abstract}


\IEEEpeerreviewmaketitle

\section{Introduction}

\IEEEPARstart{I}{n} multiple-input multiple-output (MIMO) researches and applications, there present two kinds of beamforming implementations, i.e., analog beamforming and digital beamforming. The analog beamforming helps reap diversity with low system complexity and power consumption, while digital beamforming is more flexible in that it achieves the tradeoff between diversity and spatial multiplexing. In detail, the digital beamforming could be designed to improve the signal power, cancel out the interference, minimize the mean square error, etc. This can be realized by optimizing the digital beamformer with various objective functions \cite{1023431, 4723359, 5285160}. However, the digital beamforming, which requires higher hardware costs, leads to harder challenges for practical implementations than the analog one. In multiuser case, although a fully digital beamforming is able to achieve the optimal performance in terms of system capacity, sub-optimal linear processing is widely used due to its low complexity. In future communication systems, large-scale antenna arrays are employed to improve the quality, capacity and reliability of communcations, i.e., massive MIMO \cite{5595728, 6375940, 7558167}. For massive MIMO, the system might not be able to afford the pure digital beamforming because of the growing power consumption, hardware cost and system complexity. To alleviate this issue, hybrid analog and digital beamforming was proposed to establish the tradeoff between complexity and performance \cite{6847111, 6717211, 7037444}. In detail, the hybrid beamforming consists of a high-dimensional analog beamformer and a low-dimensional digital beamformer which decreases the required number of RF chains. For multiuser massive MIMO, ideas of modified linear processing \cite{6928432, 7510972, 7947159}, were introduced to achieve highly desirable performance with further reduced system complexity. Especially, in \cite{6928432}, the spectral efficiency of hybrid beamforming based on linear digital beamforming designs is proved to asymptotically approach that of the pure digital beamforming for massive MIMO. However, as sub-optimal design without optimal performance, hybrid beamforming based on linear digital beamforming designs is not proved to outperform analog processing in all cases. Serving as circumstantial evidence, from \cite{6928432, 7510972, 7947159}, the analog beamforming approximately transferred the channel matrix into a diagonal matrix in which no interference among users may asymptotically exist. %Under this consideration, the necessity of digital processing seems to be possibly marginal.
Therefore, it is reasonable to investigate whether the analog beamforming could sometimes beat some popular linear, but not mathmatically optimal hybrid beamforming. In other words, the pure analog beamforming could be enough for some specific communication scenarios.

In this work, we assume that the number of RF chains is equal to that of users and derive the uplink achievable rates of the hybrid detection using maximal ratio combining (MRC) and zero-forcing (ZF) criterions, and the pure analog detection under Rayleigh fading channels. Furthermore, we investigate the SNR and the number of antennas thresholds between the pure analog processing and hybrid processing schemes. The obtained thresholds display the conditions when the pure analog processing is suggested. Numerical results not only verify the proposed conclusions, but also show that the superiority of the pure analog detecton and hybrid detection still holds for the downlink case and mmWave channel model.

%In this work, we first derive the achievable rates of the maximal ratio combining (MRC)-based hybrid detection and the pure analog detection. Furthermore, %based on the derived expressions of achievable rates,
%we investigate the SNR and the number of antennas thresholds between the pure analog processing and hybrid processing schemes. The obtained thresholds display the conditions when the pure analog processing is suggested. Moreover, we also provide some simulation results on zero-forcing (ZF)-based hybrid processing and show that analog detection could also sometimes beat ZF-based hybrid detection.

\section{System Model}

Consider the uplink of a massive MIMO utilizing hybrid analog and digital processing with the sub-connected structure. The system is formed by a base station (BS) equipped with an array of $M$ antennas and $K$ single-antenna users. We assume that the BS owns $N_{RF}$ RF chains. Each RF chain is connected to an exclusive set of $N$ antennas through a dedicated phase shifter where $N=\frac{M}{N_{RF}}$.

%\begin{figure}
%\centering
%\begin{minipage}[]{0.32\textwidth}
%\includegraphics[width=0.8\textwidth]{fig1.eps}
%\end{minipage}
 %\caption{Multiuser massive MIMO uplink using hybrid detection.}
%\end{figure}

We consider a flat fading channel. At BS, the received signal processed by hybrid detection can be represented as
\begin{align}
\mathbf{y}=\sqrt{p}\mathbf{WAHs}+\mathbf{WAn}\label{eq02}
\end{align}
where $\mathbf{s}=[s_1,s_2,...,s_K]^T\in\mathbb{C}^{K\times1}$ denotes the vector of symbols transmitted by all users such that $\mathbb{E}[\mathbf{s}\mathbf{s}^H]=\mathbf{I}_K$, $\mathbf{H}=[\mathbf{h}_1,\mathbf{h}_2,...,\mathbf{h}_K]\in\mathbb{C}^{M\times K}$ stands for the channel matrix between the BS and all users with $\mathbf{h}_k\sim\mathcal{CN}(\mathbf{0}_M,\mathbf{I}_M)$, and $\mathbf{n}$ refers to the additive white Gaussian noise vector with $\mathbf{n}\sim\mathcal{CN}(\mathbf{0}_K,\sigma_n^2\mathbf{I}_K)$, $\mathbf{A}=[\mathbf{a}_1,\mathbf{a}_2,...,\mathbf{a}_{N_{RF}}]^T\in\mathbb{C}^{N_{RF}\times M}$ and $\mathbf{W}=[\mathbf{w}_1,\mathbf{w}_2,...,\mathbf{w}_K]^T\in\mathbb{C}^{K\times N_{RF}}$, respectively, represent the analog and digital detection matrices. According to \eqref{eq02}, the $k$-th element of $\mathbf{y}$ is given by
\begin{align}
y_k=\sqrt{p}\mathbf{w}_k^T\mathbf{A}\mathbf{h}_ks_k+\sqrt{p}\sum_{j\neq k}\mathbf{w}_k^T\mathbf{A}\mathbf{h}_js_j+\mathbf{w}_k^T\mathbf{An}.\label{eq03}
\end{align}
%Then, the ergodic achievable uplink rate of each user is
%\begin{align}
%\bar{R}_H=\mathbb{E}\left[\log_2\left(1+\frac{p|\mathbf{w}_k^T\mathbf{A}\mathbf{h}_k|^2}{p\sum\limits_{j=1,j\neq k}^K|\mathbf{w}_k^T\mathbf{A}\mathbf{h}_j|^2+\sigma_n^2\|\mathbf{w}_k^T\mathbf{A}\|_F^2}\right)\right].\label{eq04}
%\end{align}

\section{Uplink Rate Derivations}

\subsection{Analog Processing}

Generally, analog processing is implemented by phase shifters which conduct rotations to signal phases. In this stage, the analog processing is designed by selecting the optimal angles to maximize the signal power of each user, like the designs in \cite{6928432, 7510972, 7947159}. Interference among users is left to be addressed via the following digital processing if hybrid processing applies. In this work, we take the assumption that each subarray is responsible for one user which implies that $N_{RF}=K$. Thus, for each user, the analog processing problem can be formulated as
\begin{align}
\max\limits_{\mathbf{a}_k} \quad&\|\mathbf{a}_k^T\mathbf{h}_k\|_F^2\\
{\rm s.t.}\quad &|a_{k,i}|=\left\{\begin{array}{ll}
\frac{1}{\sqrt{N}}, & N(k-1)+1\leq i\leq Nk\\
0, & {\rm otherwise}
\end{array}\right.
\end{align}
where $a_{k,i}$ is the $i$-th element of $\mathbf{a}_k$. Then, it is not difficult to get
\begin{align}
a_{k,i}=\left\{\begin{array}{ll}
\frac{1}{\sqrt{N}}\frac{h_{k,i}^*}{|h_{k,i}|}, & N(k-1)+1\leq i\leq Nk\\
0, & {\rm otherwise}
\end{array}\right.\label{eq05}
\end{align}
where $h_{k,i}$ is the $i$-th element of $\mathbf{h}_{k}$. For the $k$-th user, its effective channel after analog processing is defined as
\begin{align}
\mathbf{g}_k\triangleq\mathbf{A}\mathbf{h}_k.\label{eq06}
\end{align}
For the analog detection, the digital processing part is removed from the system in the physical aspect. In contrast, in the mathematical aspect, $\mathbf{W}$ does not cope with the multiuser interference which indicates that
\begin{align}
\mathbf{W}=\mathbf{I}_K.\label{eq7.5}
\end{align}
With the design above, we derive the uplink rate in the following theorem.
\begin{myTheo}
The uplink ergodic rate per user using the analog detection is characterized as
\begin{align}
\bar{R}_A\approx\log_2\left(1+\frac{\gamma \frac{\pi N}{4}}{\gamma(K-1)+1}\right)\label{urana}
\end{align}
where $\gamma=\frac{p}{\sigma_n}$ is the signal-to-noise ratio (SNR).
\end{myTheo}

Proof: The ergodic rate can be well approximated in massive MIMO by
\begin{align}
\bar{R}_A=&\mathbb{E}\left[\log_2\left(1+\frac{p|\mathbf{w}_k^T\mathbf{A}\mathbf{h}_k|^2}{p\sum\limits_{j\neq k}|\mathbf{w}_k^T\mathbf{A}\mathbf{h}_j|^2+\sigma_n^2\|\mathbf{w}_k^T\mathbf{A}\|_F^2}\right)\right]\nonumber\\
\overset{(a)}{\approx}&\log_2\left(1+\frac{p\mathbb{E}[|\mathbf{w}_k^T\mathbf{A}\mathbf{h}_k|^2]}{p\sum\limits_{j\neq k}\mathbb{E}[|\mathbf{w}_k^T\mathbf{A}\mathbf{h}_j|^2]+\sigma_n^2\mathbb{E}[\|\mathbf{w}_k^T\mathbf{A}\|_F^2]}\right)\nonumber\\
\overset{(b)}{=}&\log_2\left(1+\frac{\gamma\mathbb{E}\left[\left|g_{k,k}\right|^2\right]}{\gamma\sum\limits_{j\neq k}\mathbb{E}\left[\left|g_{k,j}\right|^2\right]+\mathbb{E}[\|\mathbf{w}_k^T\mathbf{A}\|_F^2]}\right)\label{Ana}
\end{align}
where (a) is achieved by applying \cite[Lemma 1]{6816003}, (b) uses \eqref{eq06} and \eqref{eq7.5}. The work of deriving the ergodic rate is now to calculate the expectation of terms in \eqref{Ana}.

First of all, we have the following term in \eqref{Ana} as
\begin{align}
\mathbb{E}[\|\mathbf{w}_k^H\mathbf{A}\|_F^2]\overset{(a)}{=}\mathbb{E}\left[\|\mathbf{w}_k^H\|_F^2\right]=1\label{eq11}
\end{align}
where (a) is due to the fact that $\mathbf{A}\mathbf{A}^H=\mathbf{I}_K$ from \eqref{eq05}. Subsequently, the elements of $\mathbf{g}_k$, $g_{k,i}$, are investigated. Recalling $\mathbf{h}_k\sim\mathcal{CN}(\mathbf{0}_M,\mathbf{I}_M)$, $\{h_{k,i}\}$'s are independent and identically distributed (i.i.d.) complex Gaussian variables with zero mean and unit variance. Thus, $|h_{k,i}|$'s follow i.i.d. Rayleigh distribution with mean $\frac{\sqrt{\pi}}{2}$ and variance $1-\frac{\pi}{4}$. Applying the %Lindeberg-L\'{e}vy
Central Limit Theorem and owing to the fact that $g_{k,k}=\frac{1}{\sqrt{N}}\sum\limits_{i=N(k-1)+1}^{Nk}|h_{k,i}|$, we get
\begin{align}
g_{k,k}\sim&\mathcal{N}\left(\frac{\sqrt{\pi N}}{2},1-\frac{\pi}{4}\right)\label{eq10}
\end{align}
for large $N$ in massive MIMO. Similarly, it is easy to get $g_{k,i}\sim\mathcal{CN}(0,1), i\neq k$ according to the Lindeberg-L\'{e}vy Central Limit Theorem. Thus, we have
\begin{align}
\mathbb{E}\left[\left(g_{k,k}\right)^2\right]=&\mathbb{E}\left[g_{k,k}\right]^2+\mathbb{V}\left[g_{k,k}\right]=\frac{\pi N}{4}+\left(1-\frac{\pi}{4}\right).\label{eqanasig}
\end{align}

Analogously, $\mathbb{E}\left[\left|g_{k,j}\right|^2\right]=1$ can be proved.

Substituting \eqref{eq11}, \eqref{eqanasig} and $\mathbb{E}\left[\left|g_{k,j}\right|^2\right]=1$ into \eqref{Ana}, it yields
\begin{align}
\bar{R}_A\approx&\log_2\left(1+\frac{\gamma[\frac{\pi N}{4}+(1-\frac{\pi}{4})]}{\gamma (K-1)+1}\right)\nonumber\\
\overset{(a)}{\rightarrow}&\log_2\left(1+\frac{\gamma \frac{\pi N}{4}}{\gamma(K-1)+1}\right)\label{anaapp}
\end{align}
where (a) utilizes $\frac{\frac{\pi N}{4}}{\frac{\pi N}{4}+1-\frac{\pi}{4}}\rightarrow1$ due to the fact that $M,N\rightarrow\infty$ in the large numbers of antennas regime with small numbers of users.

\subsection{Hybrid Processing}

For hybrid processing, if the digital processing is based on MRC, the BS calculates the detection matrix as
\begin{align}
\mathbf{W}^{\rm MRC}=\mathbf{G}^H\label{eq07}
\end{align}
where $\mathbf{G}=[\mathbf{g}_1,\mathbf{g}_2,...,\mathbf{g}_K]$. For the uplink massive MIMO, we derive a tractable expression of the achievable uplink rate which is given in the following theorem.

\begin{myTheo}
The uplink ergodic rate per user using the hybrid detection is characterized as
\begin{align}
\bar{R}_H^{\rm MRC}\approx\log_2\left(1+\frac{\gamma (\frac{\pi N}{4}+K)^2}{\gamma(K-1)(\frac{\pi N}{2}+K)+\frac{\pi N}{4}+K}\right).\label{eq08}
\end{align}
\end{myTheo}

Proof: See Appendix A.

Apart from MRC, ZF is another well-known linear receiver which is denoted as
\begin{align}
\mathbf{W}^{\rm ZF}=(\mathbf{G}^H\mathbf{G})^{-1}\mathbf{G}^H.
\end{align}
We also derive a tractable expression of the achievable uplink rate for ZF-based hybrid processing in the following proposition.

\begin{myProp}
The uplink ergodic rate per user using the hybrid detection is characterized as
\begin{align}
\bar{R}_H^{\rm ZF}\approx\log_2\left(1+\mathbb{E}[\gamma_k]\right)\label{zfth}
\end{align}
where the probability density function (p.d.f.) of $\gamma_k$ is approximated by
\begin{align}
f(\gamma_k)\approx\frac{\mathrm{exp}(-\frac{\gamma_k}{\gamma\left(\frac{\pi N}{4K}+1\right)})\left(\frac{\gamma_k}{\gamma\left(\frac{\pi N}{4K}+1\right)}\right)^{N_{RF}-K}}{\gamma\left(\frac{\pi N}{4K}+1\right)\Gamma(N_{RF}-K+1)}.
\end{align}
\end{myProp}

Proof: See Appendix B.

We take the assumption that $N_{RF}=K$ which further implies
\begin{align}
\mathbb{E}[\gamma_k]=\int_{0}^{\infty}\gamma_k\frac{\mathrm{exp}(-\frac{\gamma_k}{\gamma\left(\frac{\pi N}{4K}+1\right)})}{\gamma\left(\frac{\pi N}{4K}+1\right)}d\gamma_k=\gamma\left(\frac{\pi N}{4K}+1\right).\label{zfept}
\end{align}
Combining \eqref{zfth}-\eqref{zfept}, the ergodic uplink rate per user is denoted as
\begin{align}
\bar{R}_H^{\rm ZF}\approx\log_2\left(1+\gamma\left(\frac{\pi N}{4K}+1\right)\right).\label{urzf}
\end{align}

\section{Analog Detection vs Hybrid Detection}
To compare the performance of the pure analog processing and hybrid processing, we define the rate gap as
\begin{align}
\Delta R=&\bar{R}_H-\bar{R}_A.\label{eq21}
\end{align}

\subsection{MRC-based Hybrid Processing}

Substituting \eqref{urana} and \eqref{eq08} into \eqref{eq21}, we get
\begin{align}
\Delta R\approx&\log_2\left(1+\frac{\gamma(\frac{\pi N}{4}+K)^2}{\gamma(K-1)(\frac{\pi N}{2}+K)+\frac{\pi N}{4}+K}\right)-\log_2\left(1+\frac{\gamma\frac{\pi N}{4}}{\gamma(K-1)+1}\right).\label{eqdeltamrc}
\end{align}
Checking that $\Delta R\geq0$, we obtain $\gamma\frac{(K-1)}{K}\left(\frac{(\pi N)^2}{16}-\frac{\pi NK}{4}-K^2\right)<\left(\frac{\pi N}{4}+K\right)$. It is always checked if $\frac{(\pi N)^2}{16}-\frac{\pi NK}{4}-K^2<0$ since $\gamma\frac{K-1}{K}$ and $\frac{\pi N}{4}+K$ are always positive. By checking the quadratic inequation $\frac{(\pi N)^2}{16}-\frac{\pi NK}{4}-K^2<0$, we then get $K>\frac{\pi (\sqrt{5}-1)}{8}N$. Knowing that $N=\frac{M}{K}$, it yields
\begin{align}
K^2>\frac{\pi (\sqrt{5}-1)}{8}M\approx\frac{M}{2}.\label{eq23}
\end{align}

\begin{myOb}
\emph{If $K^2>\frac{M}{2}$, the MRC-based hybrid detection outperforms the pure analog detection whatever $\gamma$ is.}
\end{myOb}

%This observation relates to the impact of interference and noise on the system performance. Specifically, when $K$ is small, the impact of noise dominates the system performance since the multiuser interference is much less significant than noise particularly. The detection design should focus on improving the signal power but not cancelling the multiuser interference. Therefore, the pure analog processing, which is designed to maximize the signal power, behaves well enough for the system performance. On the other hand, when $K$ grows large, the interference becomes pronounced and its impact on the system performance becomes dominant. As a result, the hybrid processing using digital processing becomes better than the pure analog processing which only strenghtens the signal power via analog beamforming to combat noise.

This observation relates to the impact of interference and noise on the system performance. Specifically, when $K$ is large, the interference becomes pronounced and its impact on the system performance becomes dominant. As a result, the hybrid processing using digital processing becomes better than the pure analog processing which only strenghtens the signal power via analog beamforming to combat noise.

Otherwise, when $K^2<\frac{M}{2}$, the superiority of hybrid detection and analog detection depends on $\gamma$. It follows that
\begin{align}
\Delta R\left\{\begin{array}{ll}
\geq 0, & \gamma\leq\eta_1\\
<0, & \gamma>\eta_1
\end{array}\right.\label{eq24}
\end{align}
where $\eta_1=\frac{K\left(\frac{\pi N}{4}+K\right)}{(K-1)\left(\left(\frac{\pi N}{4}\right)^2-\frac{\pi NK}{4}-K^2\right)}$.

\begin{myOb}
\emph{For $K^2<\frac{M}{2}$, the pure analog detection outperforms the MRC-based hybrid detection if $\gamma>\eta_1$ but the MRC-based hybrid detection exceeds the pure analog detection if $\gamma<\eta_1$.}
\end{myOb}

Combining Observation 1 and 2, it is revealed that, for low SNRs, the MRC-based hybrid detection always outperforms the analog detection. While for high SNRs, the analog detection has better performance than the MRC-based hybrid detection if $K$ is small but is beaten by the MRC-based hybrid one if $K$ is large. In particlar, from \eqref{eqdeltamrc} for $\bar{R}_H,\bar{R}_A\gg 1$, it indicates
\begin{align}
\Delta R\overset{(a)}{\rightarrow}&\log_2\left(\frac{\gamma(\frac{\pi N}{4}+K)^2}{\gamma(K-1)(\frac{\pi N}{2}+2K)+\frac{\pi N}{4}+K}\right)-\log_2\left(\frac{\gamma\frac{\pi N}{4}}{\gamma(K-1)+1}\right)\nonumber\\
\overset{(b)}{\rightarrow}&\log_2\left(\frac{\gamma(\frac{\pi N}{4}+K)}{2\gamma(K-1)+2}\right)-\log_2\left(\frac{\gamma\frac{\pi N}{4}}{\gamma(K-1)+1}\right)\nonumber\\
=&\log_2\left(1+\frac{K}{\frac{\pi N}{4}}\right)-1\nonumber\\
\overset{(c)}{=}&\frac{4}{\pi\ln2}\frac{K^2}{M}-1+o\left(\frac{K}{N}\right)
\end{align}
where (a) is achieved by using $\frac{\frac{\pi N}{2}+K}{\frac{\pi N}{2}+2K}\rightarrow1$ when $N\rightarrow\infty$, (b) is obtained since $\frac{2\gamma(K-1)+1}{2\gamma(K-1)+2}\rightarrow1$ for large $\gamma$, and (c) applies the Taylor's expansion for small $\frac{K}{N}\ll1$. Since we consider a massive MIMO ($M\rightarrow\infty$) with a limited number of RF chains (small $K$), $N=\frac{M}{K}\gg K$ is always correct.

\begin{myOb}
\emph{The rate gap between the two detection schemes increases linearly with the system parameter $\frac{K^2}{M}$ in high SNR scenario.}
\end{myOb}

\subsection{ZF-based Hybrid Processing}

From \eqref{urana}, \eqref{urzf} and \eqref{eq21}, we can write
\begin{align}
\Delta R=&\log_2\left(1+\gamma\left(\frac{\pi N}{4K}+1\right)\right)-\log_2\left(1+\frac{\gamma \frac{\pi N}{4}}{\gamma(K-1)+1}\right).
\end{align}
It is not difficult to obtain
\begin{align}
\Delta R\left\{\begin{array}{ll}
\geq 0, & \gamma\geq\eta_2\\
<0, & \gamma<\eta_2
\end{array}\right.\label{eq24}
\end{align}
where $\eta_2=\frac{\frac{\pi N}{4K}(K-1)-1}{\frac{\pi N}{4K}(K-1)+K-1}$.

\begin{myOb}
\emph{The pure analog detection has better performance than the ZF-based hybrid detection if $\gamma<\eta_2$ but is beaten by the ZF-based hybrid detection if $\gamma>\eta_2$.}
\end{myOb}

\section{Simulation results}

%\begin{figure*}[!t]
%\centering
%\begin{minipage}{0.64\textwidth}
%\centering
%\subfigure[]{
%\centering
%\includegraphics[width=0.45\textwidth]{fig1a.eps}
%}
%\subfigure[]{
%\centering
%\includegraphics[width=0.45\textwidth]{fig2b.eps}
%}
%\caption{Uplink sum rates with $M=100$ in (a) and {\color{blue}$K=10$,} SNR$=10$dB in (b).}
%\end{minipage}
%\hfill
%\begin{minipage}{0.32\textwidth}
%\centering
%\includegraphics[width=0.9\textwidth]{fig4.eps}
%\caption{Uplink sum rates with $M=64$.}
%\end{minipage}
%\end{figure*}

%\begin{figure}[htbp]
%\centering
%\includegraphics[width=0.45\textwidth]{fig4.eps}
%\caption{Uplink sum rates with $M=64$.}
%\end{figure}

%\begin{figure}[htbp]
%\centering
%\includegraphics[width=0.45\textwidth]{downlink.eps}
%\caption{Downlink rates versus the SNRs with $M=64$.}
%\end{figure}

%\begin{figure}[htbp]
%\centering
%\includegraphics[width=0.45\textwidth]{mrcmmwave.eps}
%\caption{Uplink rates over mmWave channels versus the SNRs with $M=64$, $L=4$ and $\frac{d}{\lambda}=\frac{1}{2}$.}
%\end{figure}

In this section, we compare the uplink performance of the analog detection and hybrid detection based on different linear processing. Based on results in figures which display the sum rates, several observations can be made:

\begin{figure}[htbp]
%\centering
%\subfigure[]{
%\centering
\includegraphics[width=0.5\textwidth]{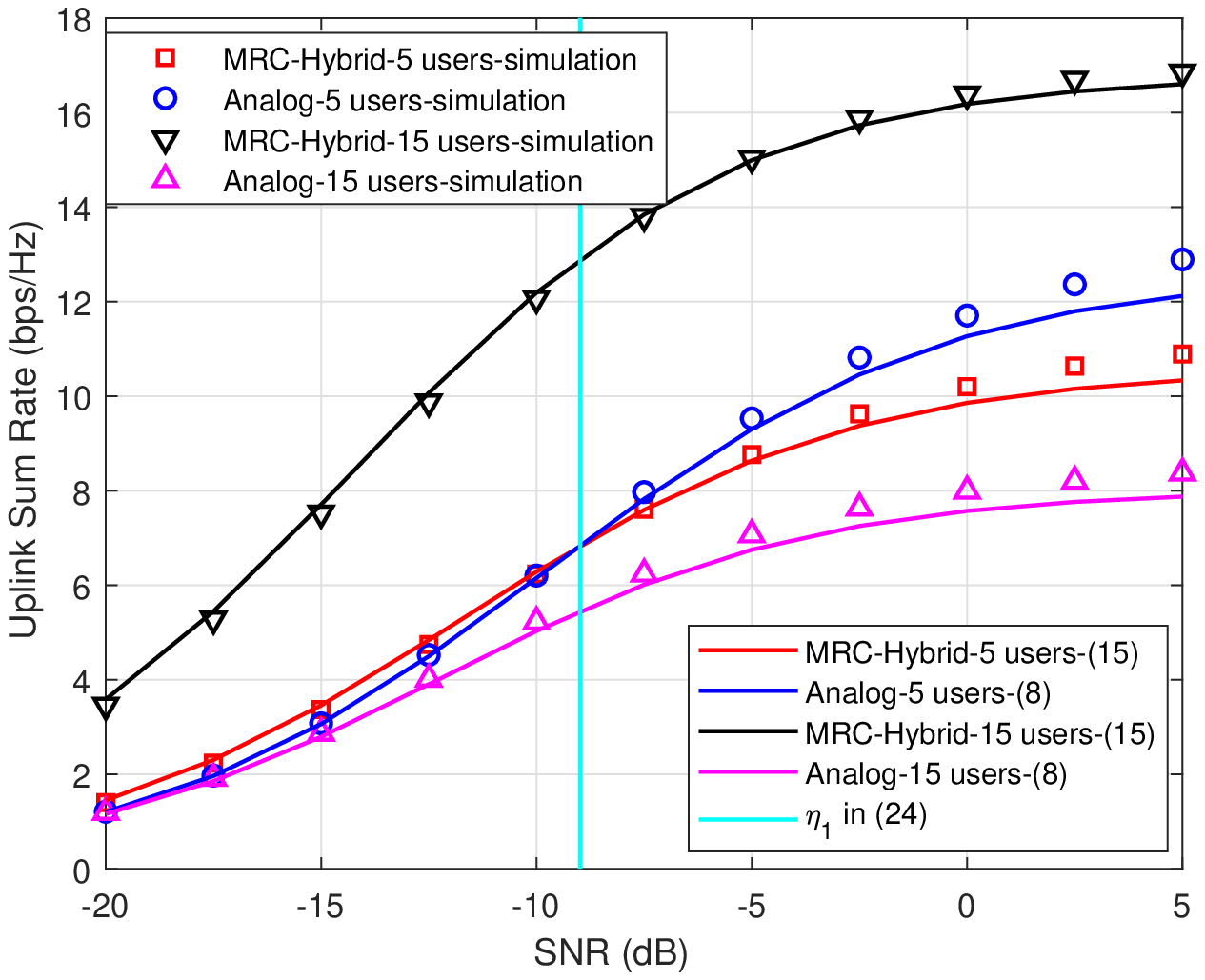}
%}
%\subfigure[]{
%\centering
\includegraphics[width=0.5\textwidth]{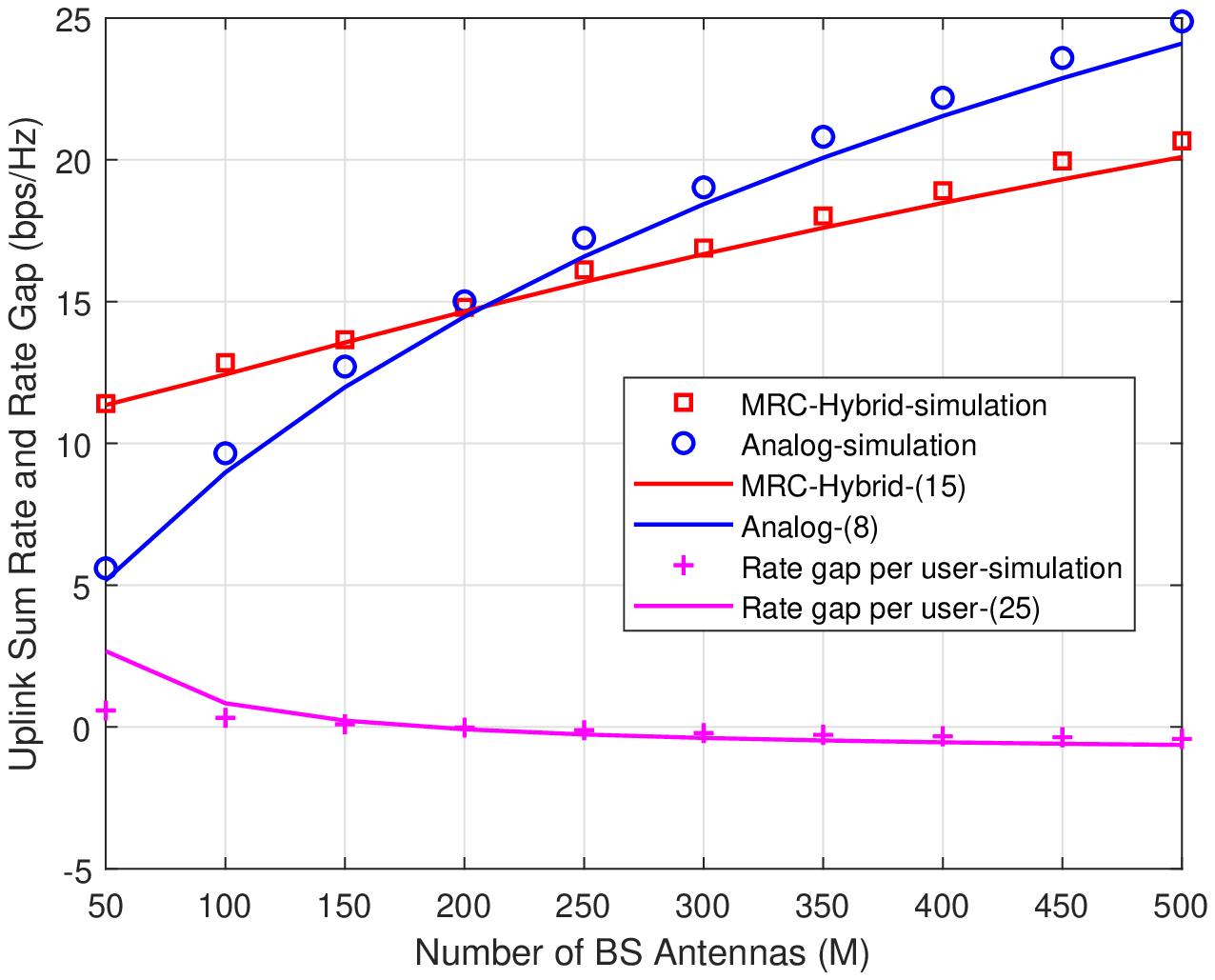}
%}
\caption{Uplink sum rates with $M=120$ in (a) and $K=10$, SNR$=10$dB in (b).}
\end{figure}

\begin{figure}[htbp]
\centering
\includegraphics[width=0.5\textwidth]{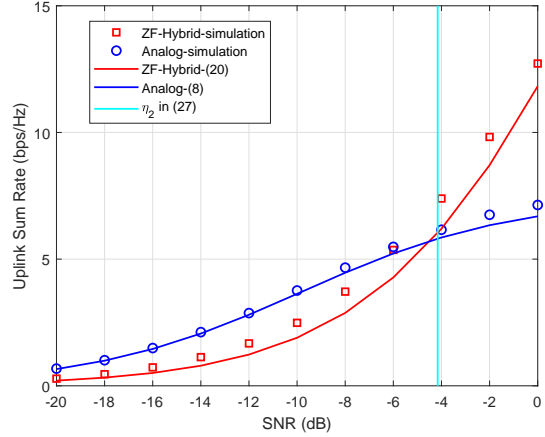}
\caption{Uplink sum rates with $M=64$ and $K=8$.}
\end{figure}

\begin{figure}[htbp]
\centering
\includegraphics[width=0.5\textwidth]{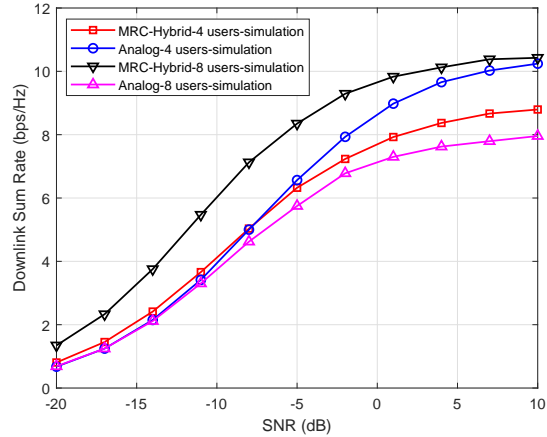}
\caption{Downlink sum rates with $M=64$.}
\end{figure}

\begin{figure}[htbp]
\centering
\includegraphics[width=0.5\textwidth]{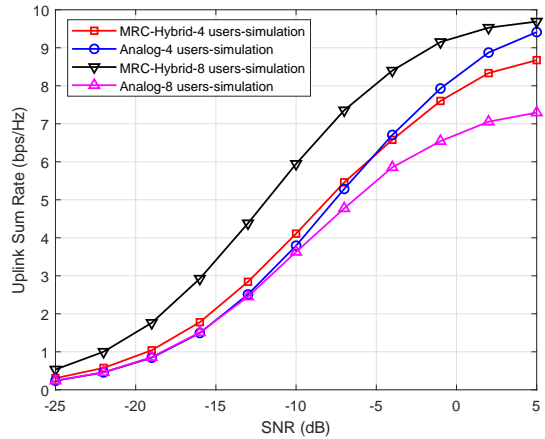}
\caption{Uplink sum rates over mmWave channels with $M=64$, $L=4$ and $\frac{d}{\lambda}=\frac{1}{2}$.}
\end{figure}

%\begin{figure*}[!t]
%\begin{minipage}{0.32\textwidth}
%\centering
%\includegraphics[width=0.9\textwidth]{fig4.eps}
%\caption{Uplink sum rates with $M=64$.}
%\end{minipage}
%\hfill
%\begin{minipage}{0.32\textwidth}
%\centering
%\includegraphics[width=0.9\textwidth]{downlink.eps}
%\caption{Downlink sum rates with $M=64$.}
%\end{minipage}
%\hfill
%\begin{minipage}{0.32\textwidth}
%\centering
%\includegraphics[width=0.9\textwidth]{mrcmmwave.eps}
%\caption{Uplink sum rates over mmWave channels with $M=64$, $L=4$ and $\frac{d}{\lambda}=\frac{1}{2}$.}
%\end{minipage}
%\end{figure*}

(1) As shown in Fig. 1 and Fig. 2, the derived expressions of achievable rates and the derived SNR thresholds, i.e., $\eta_1$ and $\eta_2$, are quite accurate for all detection methods (pure analog detection, MRC and ZF-based hybrid detection). In the derivations, we take the assumption that the system is equipped with massive MIMO ($M\rightarrow\infty$) with a limited number of RF chains (small $K$) due to practical constraints, which implies that $N=\frac{M}{K}$ can be fairly large. It is notable that the number of antennas at each subarray may not always be very large in practical applications with a finite number of BS antennas. It is discovered from Fig. 1(b) that the derived expressions are accurate even when $N$ is not too large ($M=50$, $K=10$, $N=5$). Therefore, it is still acceptable and appropriate to assume that $N$ is large in derivations.

(2) Simulation results verify the observations. In Fig. 1(a), for $K^2>\frac{M}{2}$, the MRC-based hybrid detection outperforms the analog detection whatever the SNR is; otherwise, the analog detection outperforms the MRC-based hybrid detection at high SNRs ($\gamma>\eta_1$) but is beaten by the MRC-based hybrid detection at low SNRs ($\gamma<\eta_1$). While in Fig. 1(b), it is illustrated that the hybrid detection enjoys better performance with a small $M$ but is beaten by the analog detection when $M$ is large enough. For the ZF-based hybrid processing (Fig. 2), it has worse performance than the analog processing in the low-SNR region ($\gamma<\eta_2$) but outperforms the analog processing when the SNR is large ($\gamma>\eta_2$).

(3) In Fig. 3, we provide some numerical results for the downlink channels. It is revealed that conclusions on the superiority of the pure analog detection and hybrid detection are similar with those for uplink case. More specifically, the hybrid detection does not always outperform the pure analog detection in all cases. If the number of users is relatively large ($K = 8$), the sum rate of the MRC-based hybrid processing is always larger than the pure analog detection at all SNRs. Whereas if the number of users is small ($K = 4$), the MRC-based hybrid processing outperforms the pure analog detection only when the SNR is low.

(4) Apart from Rayleigh channels, hybrid/analog processing can also be applied to mmWave communications. For the mmWave channels, the geometric channel model for user $k$ can be expressed as
\begin{align}
\mathbf{h}_k=\sqrt{\frac{M}{L}}\sum\limits_{l=1}^L\alpha_l^k\mathbf{a}_{BS}(\phi_l^k)
\end{align}
where $L$ denotes the number of propagation paths from BS to user, and $\alpha_l^k\sim\mathcal{CN}(0,1)$ represents the complex gain of the $l$-th path. Variable $\phi_l^k$ is the azimuth angle of departure (AOD) of the $l$-th path which follows uniform distribution over $[0,2\pi)$, and $\mathbf{a}_{BS}(\phi_l^k)$ is the antenna array response vector of the BS which depends on specific array structures. For uniform linear arrays (ULAs) in our simulations, $\mathbf{a}_{BS}(\phi_l^k)$ is defined as
\begin{align}
\mathbf{a}_{BS}(\phi_l^k)=\frac{1}{\sqrt{M}}\left[1,e^{j\frac{2\pi}{\lambda}d\sin(\phi_l^k)},...,e^{j(M-1)\frac{2\pi}{\lambda}d\sin(\phi_l^k)}\right]^H
\end{align}
where $\lambda$ is the signal wavelength and $d$ is the distance between adjacent antenna elements. Results in Fig. 4 show that the analog detection can sometimes outperform the hybrid one in some cases even over mmWave channels. In detail, when the number of users is small ($K=4$), the analog detection falls to beat the hybrid one at low SNRs but outperforms hybrid detection when the SNR becomes large. However, when the number of users is relatively large ($K=8$), the hybrid processing has better performance than analog processing for all SNRs.

\section{Conclusions}

We have derived the achievable uplink rates in massive MIMO for the pure analog processing and hybrid processing with the sub-connected structure under Rayleigh fading channels. It is shown that the achievable rates of the MRC and ZF-based hybrid processing can not always be larger than that of the analog processing. The SNR and the number of antennas thresholds between the pure analog detection and hybrid detection schemes.
%Particularly, in high SNR scenario, the rate gap between two detection methods is approximated as $\Delta R>-1$ for large $M$.
Simulation results verify derived conclusions not only on uplink Rayleigh fading channels but also for the downlink case and mmWave channels. Furthermore, the derivations of the downlink case is an interesting while challenging topic in our future work.

\appendices
\section{Proof of Theorem 2}
Similar with steps in \eqref{Ana}, the ergodic rate of MRC-based hybrid beamforming can be well approximated by
\begin{align}
\bar{R}_H^{MRC}\overset{(a)}{\approx}&\log_2\left(1+\frac{\gamma\mathbb{E}[|\mathbf{g}_k^H\mathbf{g}_k|^2]}{\gamma\sum\limits_{j\neq k}\mathbb{E}[|\mathbf{g}_k^H\mathbf{g}_j|^2]+\mathbb{E}[\|\mathbf{g}_k^H\mathbf{A}\|_F^2]}\right).\label{eq09}
\end{align}
where (a) uses \eqref{eq06} and \eqref{eq07}.

To complete the proof, we focus on the expectation of terms in \eqref{eq09}. Using the similar steps in \eqref{eq11}, we obtain
\begin{align}
\mathbb{E}[\|\mathbf{g}_k^H\mathbf{A}\|_F^2]=\mathbb{E}[\|\mathbf{g}_k^H\|_F^2]=(\frac{\pi N}{4}+K-\frac{\pi}{4})\label{mrc2t}
\end{align}
according to the distributions of $g_{k,k}$ and $g_{k,i}$.

We first focus on the covariance between $|g_{k,m}|^2$ and $|g_{k,n}|^2$ ($\forall m\neq n$). From \eqref{eq05} and \eqref{eq06}, it yields, e.g., for the $m$-th element in $\mathbf{g}_k$, $|g_{k,m}|^2=\frac{1}{N}\left|\sum\limits_{i=N(m-1)+1}^{Nm}h_{k,i}\frac{h_{m,i}^*}{|h_{m,i}|}\right|^2$.
Since $|g_{k,m}|^2$ and $|g_{k,n}|^2$ correspond to different elements in $\mathbf{H}$, and thanks to the assumption that elements in $\mathbf{H}$ are i.i.d. variables, they are independent which implies
\begin{align}
\mathbb{C}ov[|g_{k,m}|^2,|g_{k,n}|^2]=0.\label{eq13}
\end{align}
Then, we have that $|g_{k,i}|^2\sim\Gamma(1,1)$ for $i\neq k$, which implies
\begin{align}
\mathbb{V}\left[|g_{k,i}|^2\right]=1.\label{eq14}
\end{align}
In addition, since $\left(\frac{g_{k,k}-\frac{\sqrt{\pi N}}{2}}{\sqrt{1-\frac{\pi}{4}}}\right)^2\sim\chi^2(1)$ from \eqref{eq10}, we have
\begin{align}
\mathbb{V}\left[\left(\frac{g_{k,k}-\frac{\sqrt{\pi N}}{2}}{\sqrt{1-\frac{\pi}{4}}}\right)^2\right]=&\frac{\mathbb{V}[g_{k,k}^2-\sqrt{\pi N}g_{k,k}+\frac{\pi N}{4}]}{(1-\frac{\pi}{4})^2}\nonumber\\
=&\frac{\mathbb{V}[g_{k,k}^2]+\pi N\mathbb{V}[g_{k,k}]-2\sqrt{\pi N}\mathbb{C}ov[g_{k,k}^2,g_{k,k}]}{(1-\frac{\pi}{4})^2}\nonumber\\
=&2.\label{re2}%\nonumber\\
%=&\frac{\mathbb{V}[g_{k,k}^2]+\pi N\mathbb{V}[g_{k,k}]-2\sqrt{\pi N}\left(\mathbb{E}[g_{k,k}^3]-\mathbb{E}[g_{k,k}^2]\mathbb{E}[g_{k,k}]\right)}{(1-\frac{\pi}{4})^2}
\end{align} %$\mathbb{V}\left[\left(\frac{g_{k,k}-\frac{\sqrt{\pi N}}{2}}{\sqrt{1-\frac{\pi}{4}}}\right)^2\right]=2$.
%\begin{align}
%\mathbb{V}\left[\left(\frac{g_{k,k}-\frac{\sqrt{\pi N}}{2}}{\sqrt{1-\frac{\pi}{4}}}\right)^2\right]=2.\label{re1}
%\end{align}
%It is understandable to acquire
%\begin{align}
%&\mathbb{V}\left[\left(\frac{g_{k,k}-\frac{\sqrt{\pi N}}{2}}{\sqrt{1-\frac{\pi}{4}}}\right)^2\right]=\frac{\mathbb{V}[g_{k,k}^2-\sqrt{\pi N}g_{k,k}+\frac{\pi N}{4}]}{(1-\frac{\pi}{4})^2}\nonumber\\
%=&\frac{\mathbb{V}[g_{k,k}^2]+\pi N\mathbb{V}[g_{k,k}]-2\sqrt{\pi N}\mathbb{C}ov[g_{k,k}^2,g_{k,k}]}{(1-\frac{\pi}{4})^2}.\label{re2}%\nonumber\\
%=&\frac{\mathbb{V}[g_{k,k}^2]+\pi N\mathbb{V}[g_{k,k}]-2\sqrt{\pi N}\left(\mathbb{E}[g_{k,k}^3]-\mathbb{E}[g_{k,k}^2]\mathbb{E}[g_{k,k}]\right)}{(1-\frac{\pi}{4})^2}
%\end{align}
Due to the definition of covariance, we obtain
\begin{align}
\mathbb{C}ov[g_{k,k}^2,g_{k,k}]=&\mathbb{E}[g_{k,k}^3]-\mathbb{E}[g_{k,k}^2]\mathbb{E}[g_{k,k}]\nonumber\\
\overset{(a)}{=}&\left(\mathbb{E}[g_{k,k}]\right)^3+3\mathbb{E}[g_{k,k}]\mathbb{V}[g_{k,k}]-\mathbb{E}[g_{k,k}^2]\mathbb{E}[g_{k,k}]\nonumber\\
%=&2\mathbb{E}[g_{k,k}]\mathbb{V}[g_{k,k}]\nonumber\\
=&\sqrt{\pi N}(1-\frac{\pi}{4})\label{re3}
\end{align}
where (a) could be proved by applying $\frac{g_{k,k}-\frac{\sqrt{\pi N}}{2}}{\sqrt{1-\frac{\pi}{4}}}\sim\mathcal{N}(0,1)$ and $\mathbb{E}[x^p]=0$ for $x\sim\mathcal{N}(0,\sigma^2)$ if $p$ is odd. Combining \eqref{re2} and \eqref{re3}, it yields
\begin{align}
\mathbb{V}[g_{k,k}^2]=\pi N\left(1-\frac{\pi}{4}\right)+2\left(1-\frac{\pi}{4}\right)^2.\label{eq16}
\end{align}
Consequently, we write
\begin{align}
\mathbb{E}[|\mathbf{g}_k^H\mathbf{g}_k|^2]=&\mathbb{E}\left[\left(\sum_{i=1}^{K}|g_{k,i}|^2\right)^2\right]\nonumber\\
=&\left(\mathbb{E}\left[\sum_{i=1}^{K}|g_{k,i}|^2\right]\right)^2+\sum_{i=1}^{K}\mathbb{V}\left[|g_{k,i}|^2\right]+2\sum_{1\leq m<n\leq K}\mathbb{C}ov[|g_{k,m}|^2,|g_{k,n}|^2]\nonumber\\
\overset{(a)}{=}&\omega_1\label{eq17}
\end{align}
where (a) uses \eqref{eq13}-\eqref{eq16} and the above obtained distributions of $g_{k,k}$ and $g_{k,i}$ and $\omega_1=(\frac{\pi N}{4}+K-\frac{\pi}{4})^2+\pi N(1-\frac{\pi}{4})+2(1-\frac{\pi}{4})^2+(K-1)$.

Analogously, $\mathbb{E}[|\mathbf{g}_k^H\mathbf{g}_j|^2]=(\frac{\pi N}{2}+K-\frac{\pi}{2})\triangleq\omega_2$ can be readily proved.

Substituting \eqref{mrc2t}, \eqref{eq17} and $\mathbb{E}[|\mathbf{g}_k^H\mathbf{g}_j|^2]=(\frac{\pi N}{2}+K-\frac{\pi}{2})$ into \eqref{eq09}
\begin{align}
\bar{R}_H^{MRC}\approx&\log_2\left(1+\frac{\gamma\omega_1}{\gamma(K-1)\omega_2+(\frac{\pi N}{4}+K-\frac{\pi}{4})}\right)\nonumber\\
\overset{(a)}{\rightarrow}&\log_2\left(1+\frac{\gamma (\frac{\pi N}{4}+K)^2}{\gamma(K-1)(\frac{\pi N}{2}+K)+\frac{\pi N}{4}+K}\right)
\end{align}
where (a) uses similar steps in \eqref{anaapp}. It completes the proof.$\hfill\rule{3mm}{3mm}$

\section{Proof of Proposition 1}
Similar with steps in \eqref{Ana}, the spectral efficiency of each user is
\begin{align}
\bar{R}_H^{ZF}
=&\mathbb{E}\left[\log_2\left(1+\frac{\gamma}{[(\mathbf{G}^H\mathbf{G})^{-1}]_{k,k}}\right)\right]\nonumber\\
\approx&\log_2\left(1+\mathbb{E}\left[\frac{\gamma}{[(\mathbf{G}^H\mathbf{G})^{-1}]_{k,k}}\right]\right).%\nonumber\\
%=&\log_2\left(1+\mathbb{E}[\gamma_k]\right)
\label{eqth2}
\end{align}
%where $\gamma_k=\frac{\gamma}{[(\mathbf{G}^H\mathbf{G})^{-1}]_{k,k}}$.

Denoting $\gamma_k=\frac{\gamma}{[(\mathbf{G}^H\mathbf{G})^{-1}]_{k,k}}$, the remaining work is to calculate $\mathbb{E}[\gamma_k]$ which requires analyzing the distributions of effective channels. As mentioned in Appendix A, we have $g_{k,k}\sim\mathcal{N}\left(\frac{\sqrt{\pi N}}{2},1-\frac{\pi}{4}\right)$ and $g_{k,i}\sim\mathcal{CN}(0,1), i\neq k$. We approximate $g_{k,k}$ as $\tilde{g}_{k,k}=\frac{\sqrt{\pi N}}{2}+\alpha$ where $\alpha\sim\mathcal{CN}(0,1)$. %The approximated one, i.e., $\tilde{g}_{k,k}$ has the same first-order moments and second-order moments differing by $\frac{\pi}{4}$, as $g_{k,k}$. 
The effective channel matrix with $\tilde{g}_{k,k}$ refers to $\tilde{\mathbf{G}}$. The $i$-th row of $\tilde{\mathbf{G}}$, i.e., $\tilde{\mathbf{G}}_i$, has a complex multivariate normal distribution denoted by $\tilde{\mathbf{G}}_i\sim\mathcal{CN}(\pmb{\mu}_i,\pmb{\Sigma})$. Then $\tilde{\mathbf{G}}^H\tilde{\mathbf{G}}$ follows a complex Wishart distribution denoted by $\mathbf{Y}=\tilde{\mathbf{G}}^H\tilde{\mathbf{G}}\sim\mathcal{CW}(N_{RF},\mathbf{V},\pmb{\Sigma})$ where $\mathbf{V}=[\pmb{\mu}_1,...,\pmb{\mu}_{N_{RF}}]^T$.

A complex semi-correlated central Wishart matrix $\hat{\mathbf{Y}}\sim\mathcal{CW}(N_{RF},\hat{\pmb{\Sigma}})$, with the effective correlation matrix being $\hat{\pmb{\Sigma}}=\pmb{\Sigma}+\frac{1}{N_{RF}}\mathbf{V}^H\mathbf{V}$, has the same first-order moments and second-order moments differing by $\frac{1}{N_{RF}}\mathbf{V}^H\mathbf{V}$, as the complex non-central Wishart matrix $\mathbf{Y}$ \cite{AppWishart}. For the case of $\hat{\mathbf{Y}}\sim\mathcal{CW}(N_{RF},\hat{\pmb{\Sigma}})$, substituting p.d.f. of $\gamma_k$ into \eqref{eqth2} \cite{1023431}, the proof completes.$\hfill\rule{3mm}{3mm}$

\ifCLASSOPTIONcaptionsoff
  \newpage
\fi

\bibliographystyle{ieeetr}
\bibliography{myreference}

%It is not necessary to upload the biography when you submit your manuscript.

\end{document}